\documentclass[conference]{IEEEtran}
\IEEEoverridecommandlockouts

\usepackage{
amsmath,
graphicx,
cite,
multirow,
booktabs,
amssymb,
enumitem,
url,
amsfonts,
algorithmic,
textcomp,
xcolor,
hyperref,
}

\def\BibTeX{{\rm B\kern-.05em{\sc i\kern-.025em b}\kern-.08em
    T\kern-.1667em\lower.7ex\hbox{E}\kern-.125emX}}
    
\begin{document}

\title{SincQDR-VAD: A Noise-Robust Voice Activity Detection Framework Leveraging Learnable Filters and Ranking-Aware Optimization}

\author{
\IEEEauthorblockN{
Chien-Chun Wang\IEEEauthorrefmark{1},
En-Lun Yu\IEEEauthorrefmark{1},
Jeih-Weih Hung\IEEEauthorrefmark{2},
Shih-Chieh Huang\IEEEauthorrefmark{4},
and Berlin Chen\IEEEauthorrefmark{1}
}
\IEEEauthorblockA{
\IEEEauthorrefmark{1} 
Dept. Computer Science and Information Engineering, National Taiwan Normal University, Taiwan
}
\IEEEauthorblockA{
\IEEEauthorrefmark{2}
National Chi Nan University, Taiwan
}
\IEEEauthorblockA{
\IEEEauthorrefmark{4}
Realtek Semiconductor Corp., Taiwan
}
}

\maketitle

\begin{abstract}
Voice activity detection (VAD) is essential for speech-driven applications, but remains far from perfect in noisy and resource-limited environments. Existing methods often lack robustness to noise, and their frame-wise classification losses are only loosely coupled with the evaluation metric of VAD. To address these challenges, we propose SincQDR-VAD, a compact and robust framework that combines a Sinc-extractor front-end with a novel quadratic disparity ranking loss. The Sinc-extractor uses learnable bandpass filters to capture noise-resistant spectral features, while the ranking loss optimizes the pairwise score order between speech and non-speech frames to improve the area under the receiver operating characteristic curve (AUROC). A series of experiments conducted on representative benchmark datasets show that our framework considerably improves both AUROC and $\boldsymbol{F_2}$-Score, while using only 69\% of the parameters compared to prior arts, confirming its efficiency and practical viability.
\end{abstract}

\begin{IEEEkeywords}
Voice activity detection, signal-to-noise ratio, sinc filter, pairwise ranking loss, area under receiver operating characteristic.
\end{IEEEkeywords}

\section{Introduction}

Voice activity detection (VAD) is a fundamental task in speech processing, serving as a crucial preprocessing step for applications such as automatic speech recognition (ASR), speaker identification, speech enhancement (SE), among others \cite{sell2014,hughes2013}. The goal of VAD is to accurately distinguish speech segments from non-speech segments in an audio stream, a task that becomes particularly challenging in low signal-to-noise ratio (SNR) environments where background noise, reverberation, and other acoustic distortions obscure speech signals. With the increasing ubiquity of speech-driven applications on edge devices and resource-constrained platforms, there is a pressing need for VAD solutions that strike a balance between accuracy and computational efficiency \cite{wu2013}.

Iconic VAD methods, such as statistics-based models \cite{sohn1999}, are computationally lightweight but often perform poorly in noisy environments due to their reliance on hand-crafted features and limited adaptability. In contrast, deep learning-based VAD models have significantly improved robustness, leveraging data-driven feature extraction to achieve superior performance in complex acoustic conditions \cite{eyben2013,sainath2013, ivry2019,krishna2019,dellaferrera2020,dinkel2020,martinelli2020,lavechin2020,yoshimura2020,dinkel2021,væhrens2021,xu2021,alisamir2022,larsen2022,polvani2022,sarkar2022,sofer2022,ball2023,shi2023,wang2023,appiani2024,jia2024,karan2024,li2024,mariotte2024,yang2024}. However, many of these advancements come at a substantial cost of increased computational demands and memory footprint, thereby hindering their direct deployment on resource-limited edge devices for real-time processing \cite{hebbar2019}.

Recognizing the critical need to achieve a delicate balance between computational efficiency and detection accuracy, recent research has focused on developing lightweight VAD architectures specifically tailored for resource-constrained environments. For instance, MarbleNet \cite{jia2021} uses time-channel separable convolutions to reduce model size with minimal performance loss, while SG-VAD \cite{svirsky2023} applies stochastic gating to focus on key features for efficiency. ResectNet \cite{kopuklu2022} further refines spectral modeling by integrating a novel convolution mechanism and a frequency shift module to streamline the network architecture. More recently, TinyVAD \cite{chae2024} employs a patchify module and CSPTiny layers with grouped convolutions to achieve lightweight voice activity detection with low memory usage. While these innovative designs offer promising trade-offs, their performance often exhibits a notable decline under challenging low-SNR conditions, leaving a lot to be desired for robust VAD in high-noise ambient use cases. Additionally, the training paradigm predominantly relies on binary cross-entropy (BCE) loss \cite{de2005}, which inherently misaligns with the final evaluation metrics, creating a mismatch between optimization objectives and practical performance indicators.

\begin{figure*}[ht]
\centering
\includegraphics[width=0.98\linewidth]{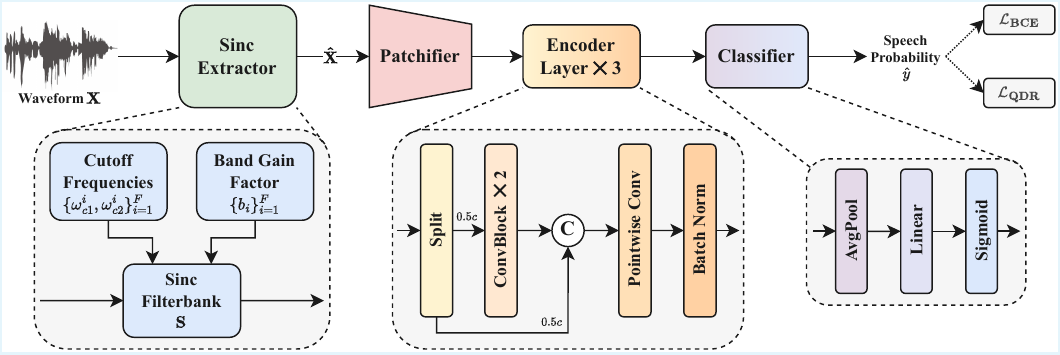}
\vspace{-5pt}
\caption{The proposed SincQDR-VAD framework consists of a feature extractor with learnable sinc filters parameterized by low/high cutoff frequencies and a band gain factor, followed by a lightweight VAD module trained with BCE and quadratic disparity ranking loss. The VAD module includes a patchify block, three encoder layers, and a classifier, with $c$ denoting the feature channel dimension.}
\label{fig:model}
\vspace{-15pt}
\end{figure*}

By clearly identifying shortcomings in existing models, our work is motivated to propose targeted improvements that enhance detection reliability and better align training with evaluation, addressing key gaps in lightweight VAD design. To this end, we put forward SincQDR-VAD, a novel VAD modeling framework meticulously designed to bolster robustness in the face of adversely noisy conditions. At the core of SincQDR-VAD lies a robust Sinc-extractor front-end, which strategically replaces conventional mel-filterbanks or standard convolutional kernels with learnable bandpass sinc filters \cite{ravanelli2018,zeghidour2018,ho2024,yu2024}. In contrast to standard convolutional layers that implicitly learn spectral features, the Sinc-extractor offers explicit and interpretable spectral control by directly modeling sub-band energy in the time domain. Each sinc filter is parameterized by learnable low- and high-cutoff frequencies and a gain factor, enabling the model to adapt its frequency resolution to better capture speech-relevant features even when temporal consistency is disrupted by noise. This front-end design proves to be particularly effective for robust speech detection in real-world low-SNR scenarios.

In addition to the architectural retrofit in the front-end, we introduce a novel ranking-aware training objective that directly aligns with VAD performance evaluation criteria. To the best of our knowledge, we are the first to formalize this notion for use on the VAD task. Specifically, we propose the quadratic disparity ranking (QDR) loss, a pairwise loss function that optimizes the area under the receiver operating characteristic curve (AUROC) \cite{zhu2022,zhang2022} by focusing on the relative ranking between speech and non-speech frames. Unlike conventional losses such as BCE, which treats each prediction independently and aims to minimize frame-wise classification error, the QDR loss encourages correct ordering by penalizing cases where a non-speech frame receives a higher score than a speech frame. This is achieved through a smooth quadratic margin, making the optimization both stable and sensitive to fine-grained ranking errors. By emphasizing pairwise consistency over absolute accuracy, our strategy offers marked robustness in noisy condition commonly encountered in real-world VAD applications. To fully exploit both global ranking performance and local classification accuracy, we adopt a hybrid loss formulation that combines the QDR loss with the traditional BCE loss. This joint objective allows the model to simultaneously refine confidence calibration and improve discrimination between speech and non-speech segments, resulting in superior performance under diverse acoustic scenarios.

To rigorously evaluate the efficacy of our proposed framework, we conduct comprehensive experiments on three benchmark datasets. SincQDR-VAD consistently outperforms several strong baselines across a range of acoustic conditions. It achieves an absolute improvement of 5\% in AUROC on the AVA-Speech dataset \cite{chaudhuri2018} and approximately 1.6\% on its noisy variant. On the ACAM dataset \cite{kim2018}, it delivers a relative improvement of 41.5\% in ${F_2}$-Score. In addition to these performance gains, SincQDR-VAD reduces the parameter count by 31\% compared to a representative lightweight VAD architecture \cite{chae2024}, demonstrating its practical suitability for deployment on resource-constrained edge devices.

In summary, the key contributions of this work are at least threefold: (1) we introduce a novel learnable front-end based on sinc filters that significantly enhances the extraction of informative features, especially under challenging noisy conditions; (2) we develop a hybrid ranking-aware training objective that effectively improves the discriminability between speech and non-speech segments, enhancing the robustness of our model to noise; and (3) we demonstrate that our proposed framework sets a new state-of-the-art when trading off between efficiency and accuracy, which seems to pave the way for scalable and reliable real-time VAD solutions for resource-limited devices operating in complex acoustic environments.

\section{Proposed Methodology}

\subsection{Framework Overview}

Fig. \ref{fig:model} depicts SincQDR-VAD, a noise-robust and lightweight VAD framework. SincQDR-VAD begins with a Sinc-extractor applying parameterized sinc filters $\mathbf{S}$ to raw waveforms $\mathbf{X}$, yielding noise-resistant features $\hat{\mathbf{X}}$. The patchify module then divides these features into patches for localized processing. The architecture includes three encoder layers that split, transform, and concatenate feature representations, capturing both fine-grained and global context. Subsequent pointwise convolution and batch normalization refine these encoded features. Finally, a classifier module uses average pooling, a linear layer, and sigmoid activation to output speech probabilities $\hat{y}$. On a separate front, the training of SincQDR-VAD combines the BCE loss for accurate classification and the QDR loss to improve ranking performance.

\subsection{Sinc-Extractor for Noise-Resilient Front End}

To address the challenges of VAD in noisy environments, we propose a task-specific Sinc-extractor that replaces the traditional mel-filterbanks commonly used in prior methods, such as TinyVAD \cite{chae2024}, with learnable sinc filters \cite{ravanelli2018}. As illustrated in Fig. \ref{fig:model}, this front-end module enhances noise robustness by extracting sub-band energy features directly from raw waveforms in the time domain. The learnable nature of these filters allows for flexible and task-adaptive feature extraction, optimizing the representation for speech detection across diverse acoustic conditions.

To compute the acoustic features, the Sinc-extractor applies a bank of parameterized sinc filters to the input waveform $x_t[n]$ at each time frame $t$. The log-energy of the $i$-th sub-band is given by
\vspace{-5pt}
\begin{equation}
\hat{x}_{t,i} = \log \left( \sum_n |x_t[n] * s_i[n]|^2 \right), i = 1, 2, \cdots, F,
\end{equation}
where $*$ denotes convolution, $F$ is the number of filters, and $s_i[n]$ is the impulse response of the $i$-th sinc filter.

The prototype of each sinc filter is parameterized by a pair of learnable cutoff frequencies, forming the difference of two sinc functions:
\vspace{-5pt}
\begin{equation}
\begin{split}
\tilde{s}_i[n] = \frac{\omega^{i}_{c2}}{\pi} \, \text{sinc} \left( \omega^{i}_{c2} n \right) -& \frac{\omega^{i}_{c1}}{\pi} \, \text{sinc} \left( \omega^{i}_{c1} n \right), \\
& \quad -\infty < n < \infty,
\end{split}
\end{equation}
where $\omega^{i}_{c1}$ and $\omega^{i}_{c2}$ are the lower and upper cutoff frequencies for the $i$-th filter, respectively, and the sinc function is defined by $\text{sinc}(k) = \sin(k) / {k}$.

This infinite-length response is delayed by $R$ samples and truncated to length $L = 2R + 1$:
\vspace{-5pt}
\begin{equation}
\hat{s}_i[n] = \tilde{s}_i[n - R], \quad 0 \leq n \leq L-1,
\end{equation}
with $\hat{s}_i[n] = 0$ elsewhere. To improve the filter expressiveness while reducing passband ripples and stopband leakage, each filter is modulated by a learnable gain parameter $b_i$ and a Hamming window $h[n]$ \cite{nuttall1981}:
\vspace{-5pt}
\begin{equation}
s_i[n] = b_i \cdot \hat{s}_i[n] \cdot h[n].
\end{equation}

The resulting sinc-filterbank $\{s_i[n]\}_{i=1}^F$ yields a feature vector $\hat{\mathbf{x}}_t = [\hat{x}_{t,1}, \cdots, \hat{x}_{t,F}]$ for each frame. The learnable filter parameters, including both cutoff frequencies $\{\omega^{i}_{c1}, \omega^{i}_{c2}\}_{i=1}^F$ and band gain factors $\{b_i\}_{i=1}^F$, are jointly optimized alongside the VAD model to promote overall task performance.

By operating directly in the time domain with optimized filters for VAD, the Sinc-extractor delivers a robust and adaptive front-end that effectively captures speech-relevant frequency components while suppressing noise. This tailored design makes it particularly suitable for VAD in challenging real-world scenarios.

\subsection{Lightweight VAD Module}

To strike a balance between accuracy and computational efficiency, we propose a lightweight VAD module tailored for real-time speech detection. Unlike previous methods like MarbleNet \cite{jia2021}, which relies on deeper convolutional stacks and higher model complexity, our design adopts a patchify operation that uses an $8 \times 8$ non-overlapping convolution kernel to split the input features $\mathbf{\hat{X}}$ into smaller patches. As shown in Fig. \ref{fig:model}, this module preserves both temporal and frequency information while significantly reducing the computational cost of subsequent processing.

The architecture is built around three encoder layers, drawing inspiration from efficient split-transform-merge methodologies \cite{devries2017,bai2018,wang2020}. Each layer features a dual-path configuration: one path focuses on local feature extraction by utilizing lightweight convolutional operations, viz. a sequence of depthwise $3 \times 3$ convolutions followed by grouped pointwise convolutions with a group size of eight, enabling efficient channel-wise processing and lower computational overhead. The other path bypasses most computations to preserve gradient flow and encourage feature reuse. These two paths are then concatenated to fuse both contextual and detailed representations, ensuring a balanced encoding of the input.

Residual connections are employed throughout the model to mitigate the vanishing gradient problem and improve feature propagation. These connections aid in preserving information from earlier layers, enabling the network to learn robust representations. On top of the encoder layers, global average pooling condenses the temporal features into a compact representation, summarizing the learned information across the time dimension. Finally, a fully connected layer followed by a sigmoid activation function produces the output $\hat{y}$, representing the probability of speech presence in the input signal $\mathbf{X}$.

\subsection{Quadratic Disparity Ranking Loss Function}

To enhance the robustness of our model under challenging acoustic conditions, we propose an innovative training strategy that emphasizes relative prediction quality rather than absolute score calibration. Central to this strategy is the quadratic disparity ranking (QDR) loss, a pairwise loss formulation designed to enhance the ability of our model in distinguishing between speech and non-speech segments by optimizing their score disparities.

In contrast to prior methods that rely solely on conventional classification losses like binary cross-entropy, which focus on minimizing individual prediction errors, the QDR loss specifically addresses the relative ordering of positive and negative samples. Specifically, it encourages the model to assign higher prediction scores to speech segments than to non-speech segments, enforcing a margin through a squared difference penalty. The QDR loss is defined by
\vspace{-3pt}
\begin{equation}\label{eq:qdr}
\mathcal{L}_{\text{QDR}} = \frac{1}{|\mathcal{P}|} \frac{1}{|\mathcal{N}|} \sum_{i \in \mathcal{P}} \sum_{j \in \mathcal{N}} \left(\max \left(0, m - \left(\hat{y}_i - \hat{y}_j\right) \right)\right)^2,
\end{equation}
where $\mathcal{P}$ and $\mathcal{N}$ denote the sets of positive (speech) and negative (non-speech) samples, respectively, $\hat{y}_i$ and $\hat{y}_j$ are their predicted scores, and $m$ represents the margin that defines the minimum desired difference between the predicted scores of positive and negative samples. This formulation penalizes cases where the score difference between speech and non-speech samples is smaller than the margin $m$, especially when a non-speech sample receives a score close to or higher than that of a speech sample. The squared term ensures smooth gradients while enhancing separation, promoting robust discrimination even in low-SNR scenarios.

Given the inherent class imbalance in voice activity detection datasets, where non-speech segments typically dominate, the pairwise nature of QDR loss improves resilience by focusing on ranking correctness instead of relying solely on the imbalanced class distribution.

To further reinforce pointwise prediction accuracy, we combine the proposed QDR loss with the standard BCE loss. The final training objective is defined by
\vspace{-5pt}
\begin{equation}\label{eq:total}
\mathcal{L}_{\text{Total}} = \lambda \mathcal{L}_{\text{QDR}} + (1 - \lambda) \mathcal{L}_{\text{BCE}},
\end{equation}
where $\lambda$ is a tunable parameter that modulates the contribution of each component. This hybrid loss encourages the model to simultaneously optimize global ranking consistency and local prediction accuracy, resulting in more robust VAD decisions across diverse and noisy environments.

\section{Experimental Setup}

\subsection{Datasets}

To train and evaluate our framework, we constructed a dataset pipeline with clean and noisy speech. For training, we used the Google Speech Commands Dataset V2 (GSC-V2) \cite{warden2018}, comprising 105,000 one-second audio clips of 35 English words from numerous speakers, providing clean, annotated speech. To simulate real-world noise, we added 2,800 environmental sound clips from Freesound (\url{https://freesound.org}) \cite{font2013}, including traffic, crowd, household, and nature sounds. This combination formed the SCF (Speech Commands + Freesound) dataset, split into training, validation, and test sets (8:1:1). We defined the central 0.2–0.83 seconds of each clip as active speech, with the remaining segments as background. For testing, a 0.15-second stride was applied to generate speech and non-speech segments, simulating realistic transitions.

For evaluation, we employed three test sets to assess the robustness of our VAD system. The first test set was derived from AVA-Speech \cite{chaudhuri2018}, consisting of 160 annotated YouTube videos categorized into clean speech, speech with noise, speech with music, and non-speech. For our binary classification task, speech categories were combined against non-speech. The second test set is a noise-corrupted version of AVA-Speech, created by mixing the original audio with environmental sounds from ESC-50 \cite{piczak2015} (2,000 clips, 50 classes) at SNRs of 10, 5, 0, -5, and -10 dB to simulate varied noise levels. The third test set is the ACAM dataset \cite{kim2018}, recorded in real-world environments (bus stop, park, construction site, indoor room) using mobile devices, with 30 minutes of audio per location.

\subsection{Configuration}

Our model operated directly in the time domain, processing raw waveform inputs sampled at 16 kHz. Each input was divided into overlapping frames of 25 ms with a 10 ms stride, capturing fine-grained temporal dynamics critical to VAD tasks. To extract meaningful acoustic representations from these waveform segments, we employed a 64-channel sinc-based filterbank ($F=64$), where the low and high cutoff frequencies of each filter are learnable parameters.

To improve generalization and noise robustness of our model, we applied data augmentation techniques tailored for time-domain processing. Specifically, we introduced random time shifts to the input waveforms with a probability of 80\%, allowing shifts ranging from -5 ms to 5 ms. Additionally, we incorporated additive white noise with an amplitude ranging from -90 dB to -46 dB, where the dB scale is defined relative to the peak amplitude of the clean waveform.

Model training was conducted over 150 epochs using stochastic gradient descent (SGD) \cite{ruder2016} with a momentum of 0.9 and a weight decay factor of 0.001 to encourage generalizable parameter updates. We set the batch size to 256. To manage learning dynamics throughout training, we adopted the WarmupHold-Decay learning rate scheduler \cite{he2019}, which allocated the first 5\% of epochs to linear warm-up, maintained a constant learning rate for the next 45\%, and applied a polynomial decay over the remaining 50\% of the training process. The margin parameter $m$ in Eq. \ref{eq:qdr} was set to 1.0 throughout all experiments. Based on empirical validation, we set the weighting coefficient $\lambda$ in Eq. \ref{eq:total} to 0.25. Our code and checkpoints are available at \url{https://github.com/JethroWangSir/SincQDR-VAD}.

\subsection{Evaluation}

To evaluate the effectiveness of the proposed  SincQDR-VAD model, we conducted a series of comparisons against representative state-of-the-art VAD models. For a fair benchmark, we followed the training and testing protocols established in the MarbleNet study \cite{jia2021}, which provides a well-established baseline for lightweight VAD systems. For post-processing, we applied a median smoothing filter using an 87.5\% overlap between adjacent segments to stabilize frame-level predictions and minimize spurious fluctuations near speech boundaries.

We primarily assessed performance using the AUROC \cite{bradley1997}, a robust metric that quantifies the trade-off between true positive rate (TPR) and false positive rate (FPR), making it particularly valuable for imbalanced or noisy conditions. Additionally, we reported the $F_2$-Score with a fixed decision threshold of 0.5 to emphasize recall and better account for false negatives, which are critical to preclude in speech applications. Finally, we analyzed the parameter count of each model, providing insights into their practicality for edge deployment where efficiency rivals accuracy in importance.

\section{Results and Discussions}

\subsection{Main Results on AVA-Speech}

The results presented in Table \ref{tab:main} clearly indicate that SincQDR-VAD surpasses several strong baselines in terms of AUROC, demonstrating its improved ability to distinguish speech from non-speech across varying acoustic conditions. In particular, SincQDR-VAD achieves a substantial gain in the $F_2$-Score, underscoring its effectiveness in reducing false negatives, which is critical in practical applications, where ignorance of speech segments would have a detrimental impact on system performance.

While SincQDR-VAD exhibits a marginally lower AUROC compared to strong baseline models such as CNN-BiLSTM \cite{wilkinson2021} and Wav2Vec2-XLS-R \cite{karan2024}, these models require much higher computational resources. In contrast, SincQDR-VAD achieves competitive accuracy with a lean architecture, making it a compelling and efficient solution for real-time VAD tasks, especially on resource-constrained platforms. Note that we did not directly compare with SG-VAD \cite{svirsky2023} because their reported results are at the utterance level, which typically yield better performance than frame-level evaluation as used in our study. Nonetheless, we have evaluated our framework at the utterance level and observed slightly better performance than SG-VAD.

\begin{table}[t]
\small
\caption{Main results on the AVA-Speech dataset.}
\vspace{-7pt}
\label{tab:main}
\centering
\setlength{\tabcolsep}{4.5pt}
\begin{tabular}{lccc}
\toprule
\bf{Model} & \bf{AUROC} & \bf{$\boldsymbol{F_2}$-Score} & \bf{Parameter (k)} \\
\toprule
CNN-TD \cite{hebbar2019} & 0.841 & - & 738 \\
ADA-VAD \cite{kim2022} & 0.853 & - & - \\
$\text{resnet\_960}$ \cite{chaudhuri2018} & 0.856 & - & 30,000 \\
MarbleNet \cite{jia2021} & 0.858 & 0.635 & 88.9 \\
TinyVAD \cite{chae2024} & 0.864 & 0.645 & 11.6 \\
ResectNet \cite{kopuklu2022} & 0.900 & - & 11.1 \\
NAS-VAD \cite{rho2022} & 0.905 & - & 151 \\
\bf{SincQDR-VAD} & \bf{0.914} & \bf{0.911} & \textbf{8.0} \\
\midrule
Braun \textit{et al.} \cite{braun2021} & 0.924 & - & 1,773 \\
CNN-BiLSTM \cite{wilkinson2021} & 0.948 & - & 552 \\
Wav2Vec2-XLS-R \cite{karan2024} & 0.962 & - & 316,000 \\
\bottomrule
\end{tabular}
\vspace{-8pt}
\end{table}

\begin{table}[t]
\small
\caption{AUROC results on the noisy variant of AVA-Speech across different SNR levels.}
\vspace{-7pt}
\label{tab:snr}
\centering
\setlength{\tabcolsep}{4.6pt}
\begin{tabular}{lcccccc}
\toprule
\multirow{2}{*}{\bf{Model}} & \multicolumn{6}{c}{\bf{SNR (dB)}} \\ \cmidrule(lr){2-7}
 & \bf{10} & \bf{5} & \bf{0} & \bf{-5} & \bf{-10} & \bf{Avg.} \\ 
\toprule
MarbleNet \cite{jia2021} & 0.838 & 0.810 & 0.765 & 0.700 & 0.620 & 0.747 \\ 
TinyVAD \cite{chae2024} & 0.859 & 0.848 & 0.823 & 0.775 & 0.691 & 0.799 \\ 
\bf{SincQDR-VAD} & \bf{0.881} & \bf{0.866} & \bf{0.836} & \bf{0.781} & \bf{0.709} & \bf{0.815} \\ 
\bottomrule
\end{tabular}
\vspace{-15pt}
\end{table}

\subsection{Results on Various SNR Levels}

Table \ref{tab:snr} presents the performance of the compared models under distinct SNR conditions. These results offer critical insight into the ability of each model to generalize in increasingly challenging acoustic environments. Across all SNR levels, SincQDR-VAD consistently outperforms baseline models, affirming its superior robustness and reliability in real-world scenarios. This improved performance in noisy settings can be attributed to the synergistic effect of the proposed Sinc-extractor front-end, which learns noise-resilient spectral features by dynamically adjusting filter frequencies and gains, together with the QDR loss, which enhances discrimination by optimizing the relative ranking between speech and non-speech frames. Our model is particularly advantageous for handling class imbalance and severe noise conditions.

The performance disparity becomes more evident at lower SNR levels, highlighting the resilience of SincQDR-VAD when background noise significantly degrades the speech signal. For example, at -10 dB, where the acoustic environment is severely contaminated and the intensity of background noise largely exceeds that of target speech, SincQDR-VAD maintains a substantial lead in AUROC. This demonstrates its ability to effectively distinguish speech even when conventional spectral and temporal cues are heavily tainted with noise. Although TinyVAD shows an advantage over MarbleNet across all SNRs, it still lags behind SincQDR-VAD by a clear margin. These findings underscore the noise robustness and superior generalization capability of SincQDR-VAD, making it especially well-suited for deployment in unpredictable and severe acoustic environments.

\begin{table}[t]
\small
\caption{AUROC (upper) and ${F_2}$-Score (lower) results on ACAM.}
\vspace{-7pt}
\label{tab:acam}
\centering
\setlength{\tabcolsep}{8pt}
\begin{tabular}{lccccc}
\toprule
\bf{Model} & \bf{Bus.} & \bf{Cons.} & \bf{Park} & \bf{Room} & \bf{Avg.} \\
\toprule
CNN-TD \cite{hebbar2019} & 0.95 & 0.77 & 0.84 & 0.95 & 0.88 \\
MarbleNet \cite{jia2021} & 0.95 & 0.78 & 0.90 & 0.96 & 0.90 \\
TinyVAD \cite{chae2024} & 0.95 & 0.94 & 0.96 & \bf{0.97} & 0.96 \\
\bf{SincQDR-VAD} & \bf{0.96} & \bf{0.98} & \bf{0.98} & 0.96 & \bf{0.97} \\
\midrule
CNN-TD \cite{hebbar2019} & 0.34 & 0.32 & 0.10 & 0.87 & 0.41 \\
MarbleNet \cite{jia2021} & 0.25 & 0.37 & 0.21 & \bf{0.94} & 0.44 \\
TinyVAD \cite{chae2024} & 0.45 & 0.73 & 0.49 & 0.93 & 0.65 \\
\bf{SincQDR-VAD} & \bf{0.92} & \bf{0.91} & \bf{0.94} & 0.92 & \bf{0.92} \\
\bottomrule
\end{tabular}
\vspace{-8pt}
\end{table}

\begin{table}[t]
\small
\caption{Ablation studies on AVA-Speech and its noisy variant.}
\vspace{-7pt}
\label{tab:ablation}
\centering
\setlength{\tabcolsep}{4.5pt}
\begin{tabular}{lcccc}
\toprule
\multirow{2}{*}{\bf{Model}} & \multicolumn{2}{c}{\bf{AVA}} & \multicolumn{2}{c}{\bf{Noisy Variant (Avg.)}} \\ \cmidrule(lr){2-3} \cmidrule(lr){4-5}
 & \bf{AUROC} & \bf{$\boldsymbol{F_2}$-Score} & \bf{AUROC} & \bf{$\boldsymbol{F_2}$-Score} \\
\toprule
\bf{SincQDR-VAD} & \bf{0.914} & \bf{0.911} & \bf{0.815} & \bf{0.864} \\
w/o Sinc-extractor & 0.889 & 0.881 & 0.784 & 0.842 \\
w/o $\mathcal{L}_{\text{QDR}}$ & 0.872 & 0.883 & 0.739 & 0.859 \\
\bottomrule
\end{tabular}
\vspace{-15pt}
\end{table}

\subsection{Performance Comparison on ACAM}

As shown in Table \ref{tab:acam}, the proposed SincQDR-VAD framework consistently demonstrates better performance across all acoustic scenarios in ACAM. In contrast to the strong baselines, it produces more balanced and resilient results, excelling in both overall discrimination and recall-oriented evaluation. While these baseline models often exhibit strengths limited to specific environments or metrics, SincQDR-VAD maintains high performance across all conditions, underscoring its ability to generalize effectively in diverse and acoustically challenging settings. These results indicate that our model not only captures the nuanced characteristics of speech activity, but also adapts gracefully to varying background conditions, making it a robust candidate for real-world deployment.

\subsection{Ablation Studies}

We conducted ablation studies to better understand the contributions of each component of the proposed SincQDR-VAD model, as summarized in Table \ref{tab:ablation}. Specifically, we evaluated three configurations: the full SincQDR-VAD model, a variant without the Sinc-extractor (w/o Sinc-extractor), and a variant excluding the QDR loss (w/o $\mathcal{L}_{\text{QDR}}$).

The full model achieves the highest AUROC and $F_2$-Score, confirming that both the Sinc-extractor and the QDR loss contribute meaningfully to overall performance. Removing the Sinc-extractor results in a moderate decline in both metrics, highlighting the role of the front-end extractor in capturing noise-robust features critical to distinguishing speech. Excluding the QDR loss causes a more significant performance drop, demonstrating its central role in enhancing the discriminative power of the proposed ranking-optimization training strategy.

\begin{figure}[t]
\centering
\includegraphics[width=0.95\linewidth]{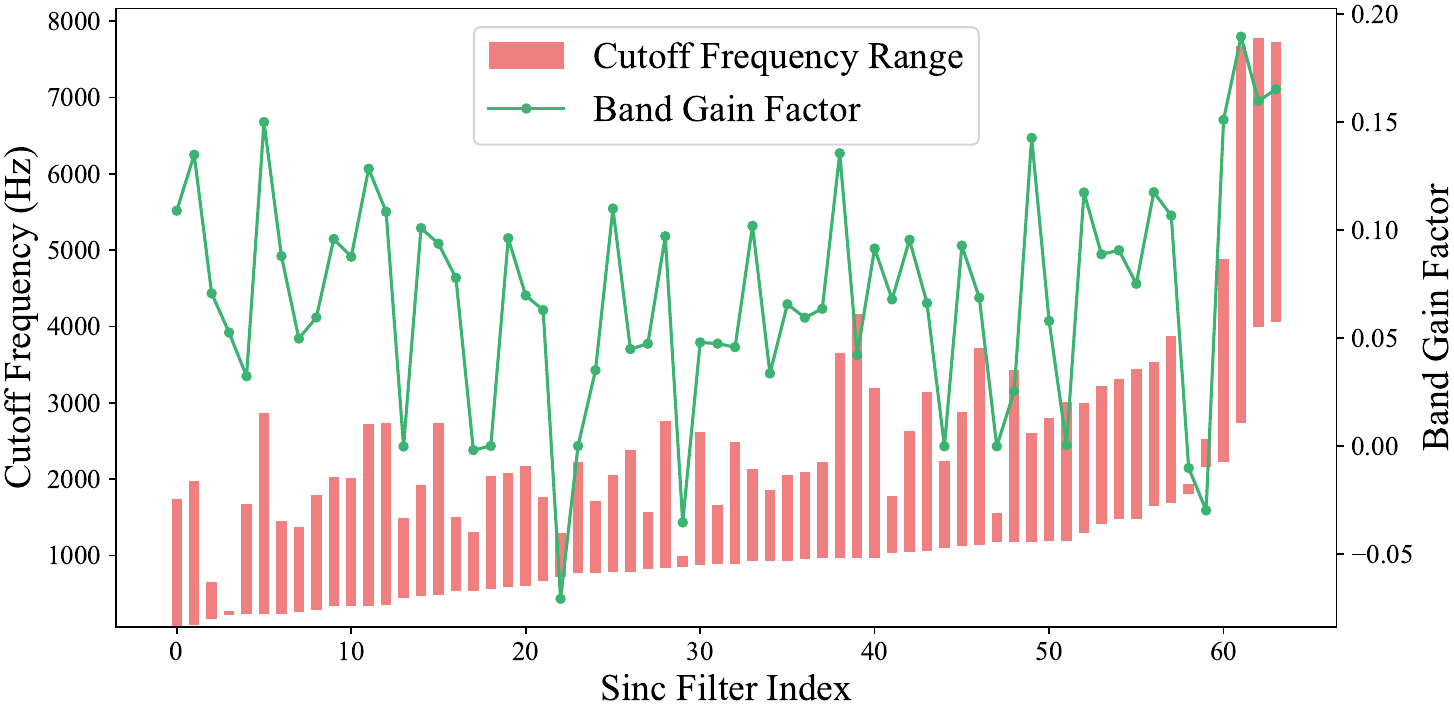}
\vspace{-10pt}
\caption{Visualization of the learned cutoff frequency ranges and corresponding band gain factors for the 64 filters of our Sinc-extractor.}
\label{fig:sinc_params}
\vspace{-5pt}
\end{figure}

\begin{figure}[t]
\centering
\includegraphics[width=0.95\linewidth]{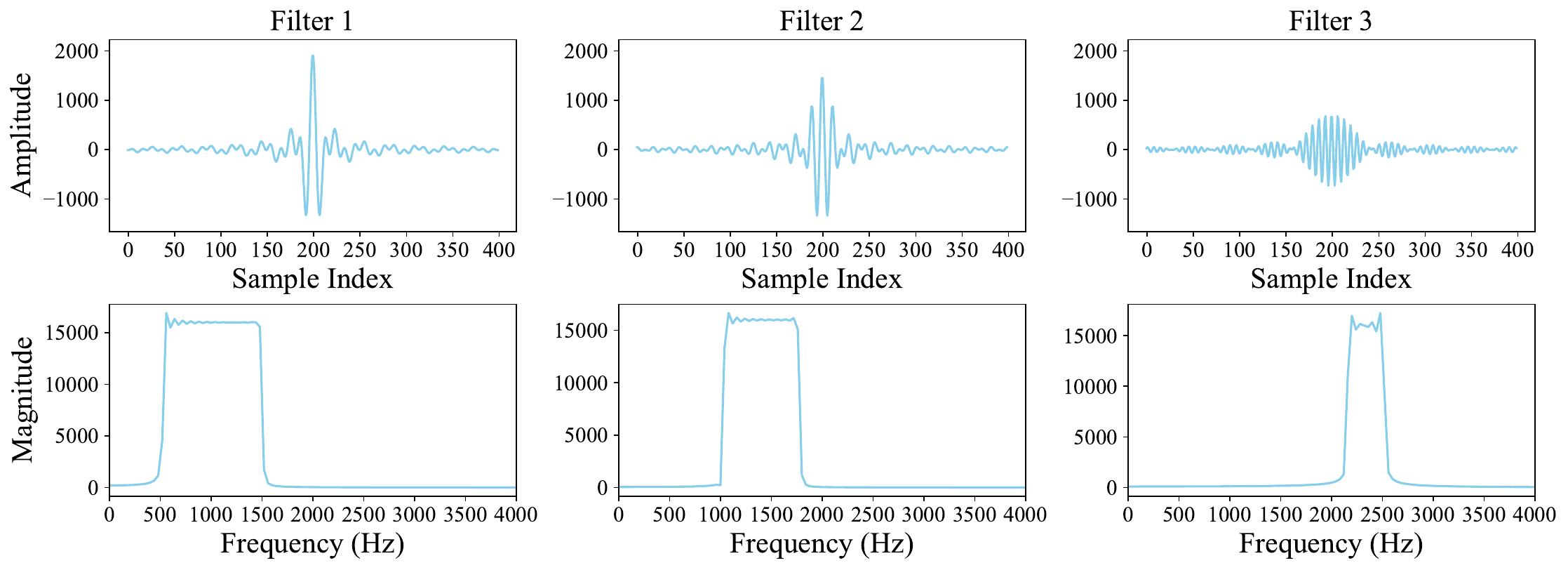}
\vspace{-10pt}
\caption{Time and frequency domain responses of three representative filters learned by the proposed Sinc-extractor. The top row shows the filters in the time domain, and the bottom row shows their magnitude frequency response.}
\label{fig:sinc_filters}
\vspace{-10pt}
\end{figure}

\subsection{Analysis of Learned Sinc Filter Characteristics}

Fig. \ref{fig:sinc_params} demonstrates that, unlike fixed mel-filterbanks with uniform frequency spacing, the learned sinc filters exhibit a frequency distribution optimized for VAD tasks. The varying cutoff frequencies (red bars) across the 64 filters indicate that the Sinc-extractor learns to focus on specific frequency regions that are most relevant for identifying speech activity. Additionally, the learned band gain factors (green line) show significant variation, suggesting that the model adaptively emphasizes critical frequency components while suppressing less informative ones, contributing to noise robustness.

Fig. \ref{fig:sinc_filters} illustrates three representative learned sinc filters in the time domain (top row) and their frequency responses (bottom row). The time-domain waveforms correspond to the impulse responses of the filters. In the frequency domain, the sharp bandpass responses highlight the ability of the filters to isolate specific frequency bands and attenuate others. This precise filtering, guided by the learned cutoff frequencies and gains (as shown in Fig. \ref{fig:sinc_params}), enables the Sinc-extractor to focus on task-relevant spectral features while mitigating out-of-band noise, thereby enhancing noise robustness. The adaptive nature of the Sinc-extractor results in learned filter characteristics, across both domains, providing a more tailored audio front-end compared to fixed mel-filterbanks. This adaptation likely improves performance in distinguishing speech from non-speech, particularly in noisy conditions.

\subsection{Visualization of Model Predictions}

\begin{figure}[t]
\centering
\includegraphics[width=0.95\linewidth]{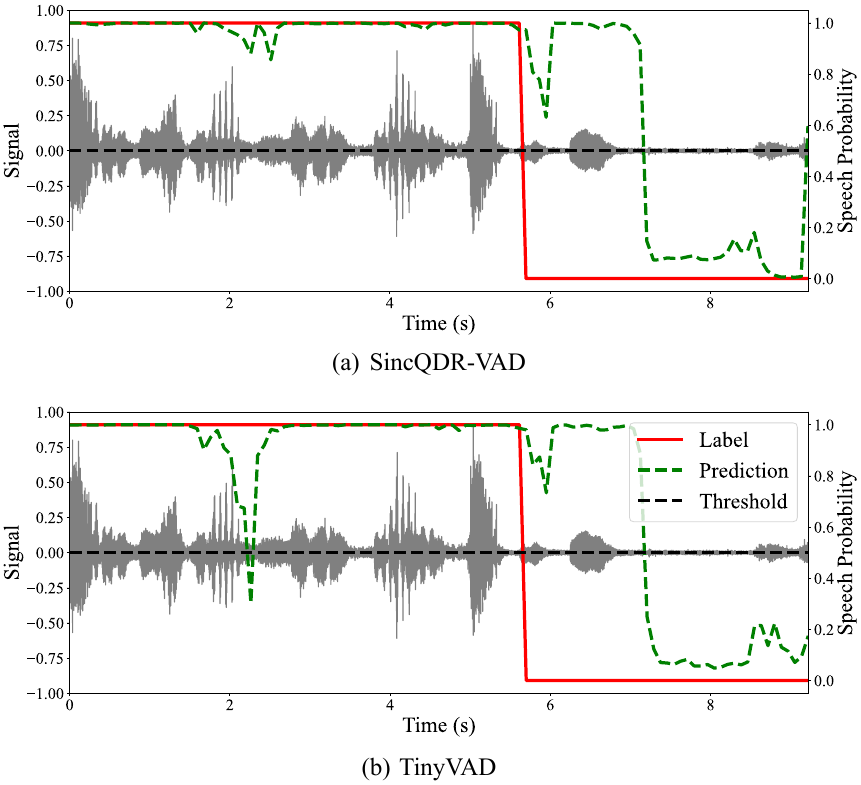}
\vspace{-5pt}
\caption{Output plots of SincQDR-VAD and TinyVAD on the noisy variant of AVA-Speech at an SNR of 0 dB.}
\label{fig:prediction}
\vspace{-10pt}
\end{figure}

Fig. \ref{fig:prediction} compares the predictions of SincQDR-VAD and TinyVAD on the noise-corrupted variant of the AVA-Speech dataset at 0 dB SNR. As illustrated, SincQDR-VAD consistently yields a smoother and more stable prediction curve, closely aligning with the ground truth. This demonstrates its robustness in accurately detecting speech amidst background noise and its enhanced generalization across challenging noisy conditions, leading to fewer false negatives and more reliable detection performance. In contrast, the prediction curve of TinyVAD is more erratic, with significant deviations from the ground truth, resulting in a higher incidence of false negatives and highlighting its greater difficulty in noisy environments. While acknowledging that both methods still exhibit false alarms requiring future work, the visualization clearly highlights the superior capability of SincQDR-VAD to handle noise and maintain stable detection.

\section{Conclusion and Future Work}

This study has proposed SincQDR-VAD, a noise-robust VAD framework that incorporates a novel Sinc-extractor front-end and a quadratic disparity-ranking loss. SincQDR-VAD demonstrates marked performance on several benchmarks, particularly in challenging low-SNR environments, while achieving an optimal balance of accuracy and efficiency suitable for real-time applications on resource-limited devices.

Future work will focus on enhancing temporal modeling and noise resilience by exploring efficient neural components such as Mamba blocks \cite{gu2023} instead of convolution. The efficiency of Mamba in long-range dependency modeling shows promise for audio sequences, potentially improving temporal context and robustness against noise, which could implicitly reduce transient false alarms. Additionally, we will address remaining false alarms from complex non-speech in dynamic environments, possibly by refining the training objective to penalize false positives more or by using richer features to better differentiate speech from challenging non-speech events. These directions are anticipated to further improve the accuracy and robustness of SincQDR-VAD in arduous real-world use cases.

\section{Acknowledgments}

This work was supported in part by Realtek Semiconductor Corporation under Grant Number 113KK01103. Any findings and implications in the paper do not necessarily reflect those of the sponsors.

\bibliographystyle{IEEEtran}
\bibliography{references.bib}

\end{document}